\documentclass[12pt]{article}\usepackage[hyperfootnotes=false]{hyperref}
\usepackage{epsfig}
\usepackage{float}
\usepackage{empheq}
\usepackage{bbold}
\usepackage{showlabels}

\usepackage[utf8]{inputenc}
\usepackage{amsmath}

\usepackage{caption}

\usepackage{amsmath}
\usepackage{amssymb}
\usepackage{graphicx}
\newlength{\abstractwidth}
\renewcommand{\thefootnote}{\fnsymbol{footnote}}
\renewcommand{\thanks}[1]{\footnote{#1}} 
\newcommand{\starttext}{
\setcounter{footnote}{0}
\renewcommand{\thefootnote}{\arabic{footnote}}}

\newcommand{\be}{\begin{equation}}
\newcommand{\bea}{\begin{eqnarray}}
\newcommand{\eea}{\end{eqnarray}}
\newcommand{\beq}{\begin{equation}}
\newcommand{\ee}{\end{equation}}

\newcommand*\widefbox[1]{\fbox{\hspace{2em}#1\hspace{2em}}}

\def\eq{&=&}

\def\simleq{\; \raise0.3ex\hbox{$<$\kern-0.75em
\raise-1.1ex\hbox{$\sim$}}\; }
\def\simgeq{\; \raise0.3ex\hbox{$>$\kern-0.75em
\raise-1.1ex\hbox{$\sim$}}\; }

\def\bi{\begin{itemize}}
\def\ei{\end{itemize}}
\def\S{Schwarzschild}

\def\dof{degrees of freedom }

\def\CC{{\cal{C}}}

\def\CI{{\cal{I}}}
\def\CJ{{\cal{J}}}

\def\CT{{\cal{T}}}

\def\bsub{ \begin{subequations}
\begin{empheq}[box=\widefbox]{align}  }
\def\esub{ \end{empheq}
\end{subequations}}

\def\1{\(  \mathbb{1} \)}

  \def\kl{k-local}

 \def\lf{\left(}
    \def\rg{\right)}

  \def\bn{\bigskip \noindent}

 \def\bm{\begin{bmatrix}}
 \def\em{\end{bmatrix}}

\makeatletter
\g@addto@macro\normalsize{%
  \setlength\abovedisplayskip{10pt}
  \setlength\belowdisplayskip{20pt}
  \setlength\abovedisplayshortskip{10pt}
  \setlength\belowdisplayshortskip{20pt}
}
\makeatother

\usepackage{color}

\begin{document}


  
\begin{titlepage}

\rightline{}
\bigskip
\bigskip\bigskip\bigskip\bigskip
\bigskip

\centerline{\Large \bf {Entanglement and Chaos in De Sitter Holography:
 }}

\bn

\centerline{\Large \bf {An SYK Example
 }}

\bigskip
\begin{center}
\bf      Leonard Susskind  \rm

\bigskip
Stanford Institute for Theoretical Physics and Department of Physics, \\
Stanford University,
Stanford, CA 94305-4060, USA \\

\bn

and Google, Mountain View, CA

\end{center}

\bn

\begin{abstract}

Entanglement, chaos, and complexity are as important for de Sitter space as for  AdS, and for black holes.  There are similarities and also great differences between AdS and dS in how these concepts are manifested in the space-time geometry.
In the first part of this paper the Ryu–Takayanagi prescription, the theory of fast-scrambling, and the holographic complexity correspondence are  reformulated for de Sitter space. Criteria are proposed for a holographic model to describe de Sitter space. The criteria can be summarized by the requirement that scrambling and complexity growth must be ``hyperfast."
In the later part of the paper I show that a certain  limit of the SYK model satisfies the hyperfast  criterion. This leads  to 
the radical conjecture that  a  limit of  SYK is indeed a  concrete, computable, holographic model  of de Sitter space. Calculations are described which support the conjecture.

\end{abstract}

\end{titlepage}

\starttext \baselineskip=17.63pt \setcounter{footnote}{0}

\Large

\tableofcontents


\section{Introduction}
There are holographic representations of AdS that are so precisely
defined that in principle one could program a quantum computer
to simulate and test them.
 By contrast there is  no such completely concrete\footnote{By concrete I mean that it is sufficiently definite that it could be programmed into a quantum computer.}  representation of de Sitter space. Not only do we not have an example; we don't even know the rules.  My purpose in this  paper is to briefly  lay out some principles, and then to construct a well-defined example that realizes those principles.

The framework  is static-patch (SP) holography  \cite{Susskind:2021omt}\cite{Susskind:2021dfc}. The assumption is  that there exists a unitary Hamiltonian quantum mechanics of a de Sitter static-patch in which the degrees of freedom are located on the stretched horizon. We  begin by considering   the roles   of  entanglement,  chaos, and complexity, and derive  necessary requirements---very different from those for AdS---for a quantum system to be dual to de Sitter space. We will see that these requirements are met by a non-standard, but  perfectly definite  limit of the  SYK system. Quantum-simulating the system should be no harder than simulating SYK in the usual range of parameters.

\subsection{Four Conjectures}

The paper revolves around four conjectures:
\begin{enumerate}
\item There exists a de Sitter generalization (which I will describe) of the RT and HRT equations for entanglement entropy in which the AdS boundary is replaced by the boundary of the static patch; namely the horizon. 
\item{The fundamental holographic horizon degrees of freedom  are ``hyperfast" scramblers which scramble on a time scale of order the dS horizon scale $R,$ rather than the fast-scrambling time scale $R \log S.$}
If true this implies that the Hamiltonian is not of the usual \kl \ type but is more complex.
\item The hyperfast scrambling property implies that complexity growth is also hyperfast. I argue that hyperfast complexity growth is the holographic dual of the most  essential feature of de Sitter space: its exponential growth.
\item The last section of the paper introduces a  ``hyperfast"  limit of the SYK model which has  the required features for a holographic de Sitter dual. This leads  to the unexpected  conjecture that a limit of SYK describes  de Sitter space.

\end{enumerate}

There are very few equations in this paper, but lots of figures. The figures efficiently  summarize  calculations that were done in earlier papers by many people.

\subsection{A Word About Time}

The metric of the static patch of de Sitter space is,
\be  
ds^2 =-f(r) dt^2 + \frac{1}{f(r)}dr^2 + r^2 d\Omega^2
\ee
where,
\be 
f(r)  = \lf 1- \frac{r^2}{R^2}  \rg. 
\ee

In what follows we will sometimes use a dimensionless  time coordinate on the stretched horizon denoted by $\omega.$ The relation between $t$ and $\omega$ is,
\be  
\omega = t/R
\ee

\begin{figure}[H]
\begin{center}
\includegraphics[scale=.6]{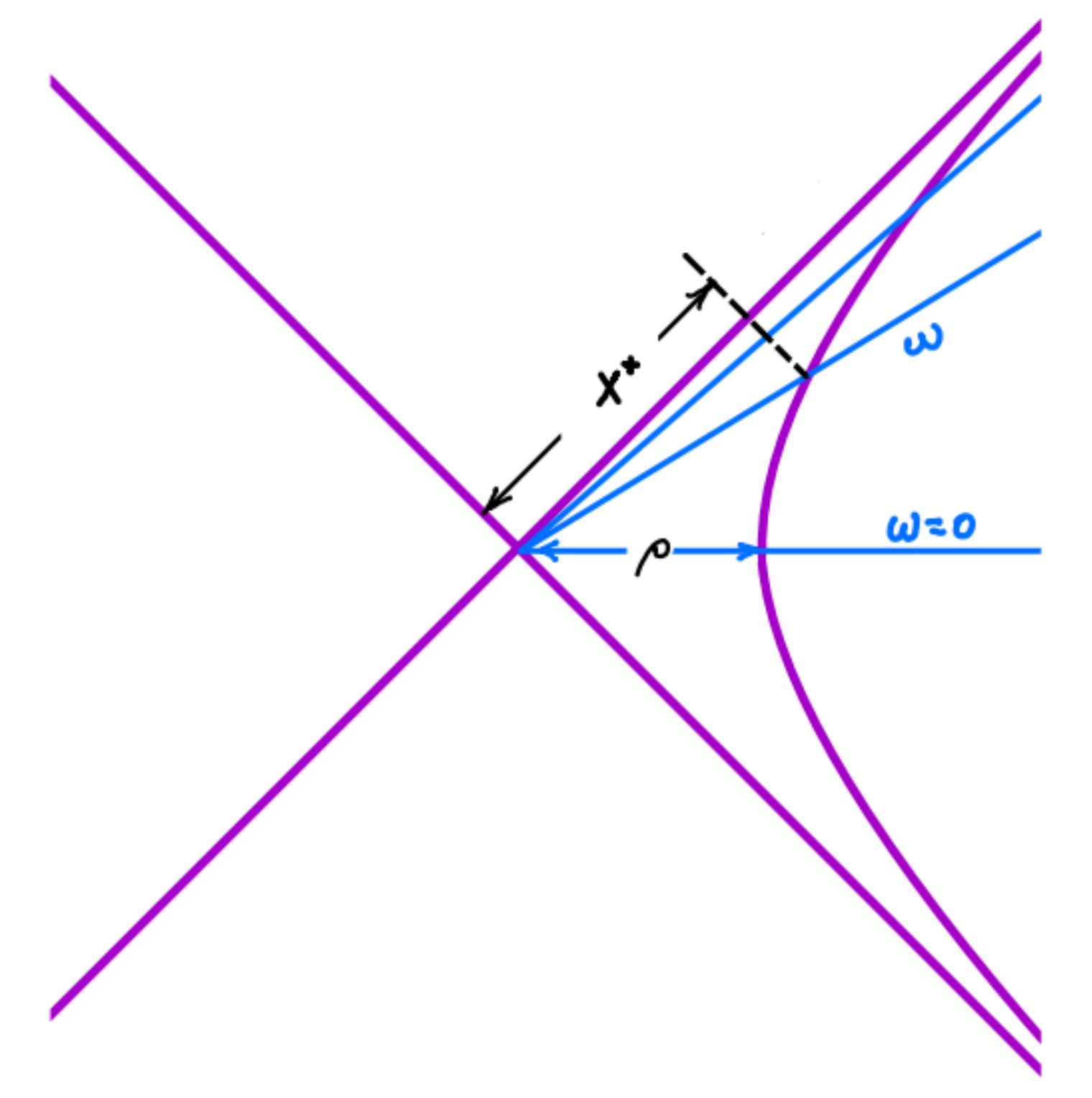}
\caption{The near horizon region of de Sitter space is approximately Rindler space. The Rindler time $\omega$ is a hyperbolic angle. The stretched horizon is the hyperbola at a distance $\rho$ from the bifurcate horizon.}
\label{time}
\end{center}
\end{figure}

Near the horizon the geometry can be approximated by Rindler space. In figure \ref{time} the relation between $\omega$ and the light-like coordinate $x^+$ is shown. The stretched horizon is indicated by the hyperbola at a distance $\rho$ from the light-cone.

The relation between $\omega$ and $x^+ $ is,
\be  
\omega = \log{x^+} - \log{\rho}
\ee
The value of $\omega$ on the stretched horizon for a given $x^+$ is mildly dependent on the value of $\rho$ that we assign to the stretched horizon. More importantly,  time differences along the stretched horizon are not dependent on $\rho.$

\bn

There is of course something uncomfortable about assuming a universal eternal time in de Sitter space. In AdS boundary time can have unlimited accuracy provided by clocks located on, or even beyond, the asymptotically frozen boundary. In the  static de sitter patch the best we can do is to assume the horizon itself defines a clock (for example through the growth of complexity), but no clock built out of a system of entropy $S$ can keep time for times longer than $e^S.$  It is also likely that no de Sitter vacuum can be stable against decay for longer than $e^S.$ The things I will talk about here do not 
require such long times.
 
 \subsection{A Word About ``Tomperature"} \label{tom}
 In a system with a finite Hilbert space the temperature can be infinite while the energy and all physical  time scales remain finite. An example is a spin in a magnetic field. The limit of infinite temperature is just the limit in which spin-up and spin-down  are equally probable, but both spin states have finite energy and the  Larmor frequency is finite. In this type of system the idea of infinite temperature can be misleading.
 
The SYK system is an example of this kind in which the temperature can be infinite without the energy-per-degree-of-freedom  being infinite or time scales going to zero. For example the energy of a fermion excitation and the decay time for the fermion two-point function  are both finite in the infinite temperature limit. 

It can be useful to define a quantity $\CT$ whch Henry Lin and I called tomperature, which reflects the finiteness of energies and time scales for such systems. The precise definition of tomperature is not important but one property involves the decay of perturbations. We will assume that tomperature is defined so that the decay of typical two-point functions behaves like $$e^{-\CT t.}$$

The important thing about $\CT$ is that it remains finite as $T\to \infty,$ and at low temperatures it is equal to the usual temperature $T.$

\section{Entanglement in de Sitter Space}
It is all but certain that the holographic principle  \cite{tHooft:1993dmi}\cite{Susskind:1994vu}, entanglement \cite{VanRaamsdonk:2010pw}\cite{Ryu:2006bv}\cite{Hubeny:2007xt}, and complexity growth \cite{Susskind:2014rva} are the essential quantum-mechanical  mechanisms that lead,  in appropriate limits,
 to the emergence of classical space-time geometry.  It would be very surprising if they did not play an outsize role in cosmology, but so far we know very little about how they work in de Sitter space, or in  other cosmological space-times.
\begin{figure}[H]
\begin{center}
\includegraphics[scale=.4]{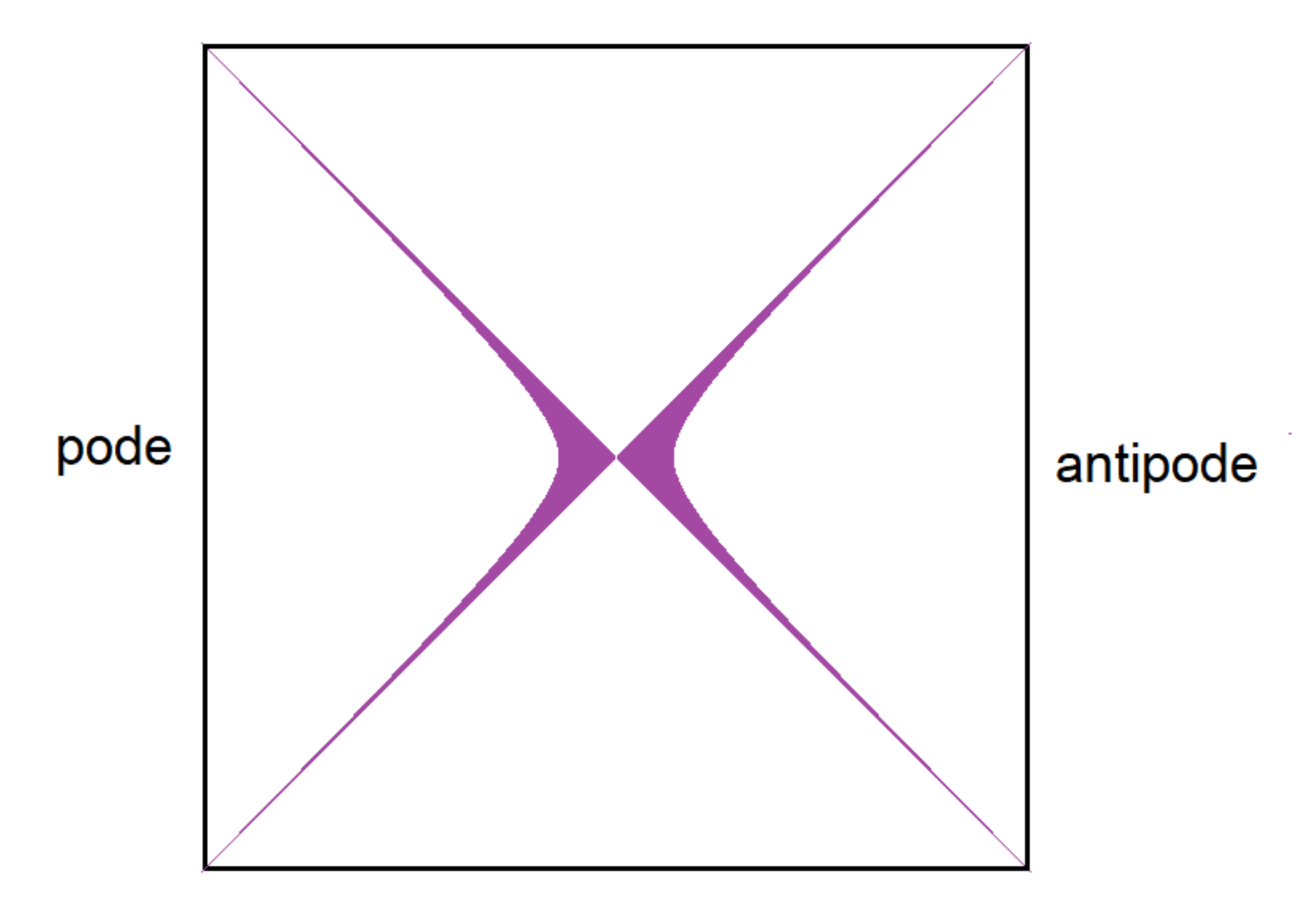}
\caption{Penrose diagram for dS and stretched horizon}
\label{stretch}
\end{center}
\end{figure}

In this paper I will assume that there is  a holographic description of a static (SP) patch in  four-dimensional  de Sitter space\footnote{The various mechanisms and calculations described in this paper apply to four dimensions except for in the last section where a possible relation between SYK and two-dimensional de Sitter space is conjectured. Generalization to other number of dimensions is non-trivial and I will not undertake the task here. }; but unlike AdS, the spatial slices of de Sitter space have no asymptotic  cold boundaries where the holographic  degrees of freedom are located. Instead, they are nominally located on the boundary of the SP (see for example \cite{Dyson:2002pf}\cite{Banks:2006rx}\cite{Susskind:2011ap}\cite{Banks:2016taq}\cite{Banks:2020zcr}\cite{Susskind:2021omt}\cite{Susskind:2021dfc}); that is to say, the stretched horizon. 

Static patches come in opposing pairs. To account for the pair,  two sets of degrees of freedom are required. The Penrose diagram of dS in figure \ref{stretch} shows such a pair of SPs along with their stretched horizons.  Following \cite{Susskind:2021omt}  the center of the SPs (sometimes thought of as the points where  observers are located) will be called the ``pode" and the ``antipode."  I'll also refer to the two SPs as the pode-patch and the antipode-patch.

\subsection{RT in dS}

Although it is  clear from the Penrose diagram that the two SPs are entangled, it would be good to have a generalization of the RT and HRT equations 
\cite{Ryu:2006bv}\cite{Hubeny:2007xt} to support this claim. Let's recall the original statement of the RT formula for AdS. It begins by dividing the boundary (on a fixed time-slice) into two subsets $A$ and $B.$

\bn 
\it The entanglement entropy of $A$ and $B$  is $1/4G$ times the minimal area
of a surface homologous to either $A$ or $B$.
\rm

\bn
In dS there is no boundary of a global time-slice but  there is a boundary to a SP, namely the horizon. Therefore we try the following formulation:


\bn 
\it The entanglement entropy between the pode and antipode patches is $1/4G$ times the minimal area
of a surface homologous to the   stretched horizon (of either side).
\rm

\begin{figure}[H]
\begin{center}
\includegraphics[scale=.3]{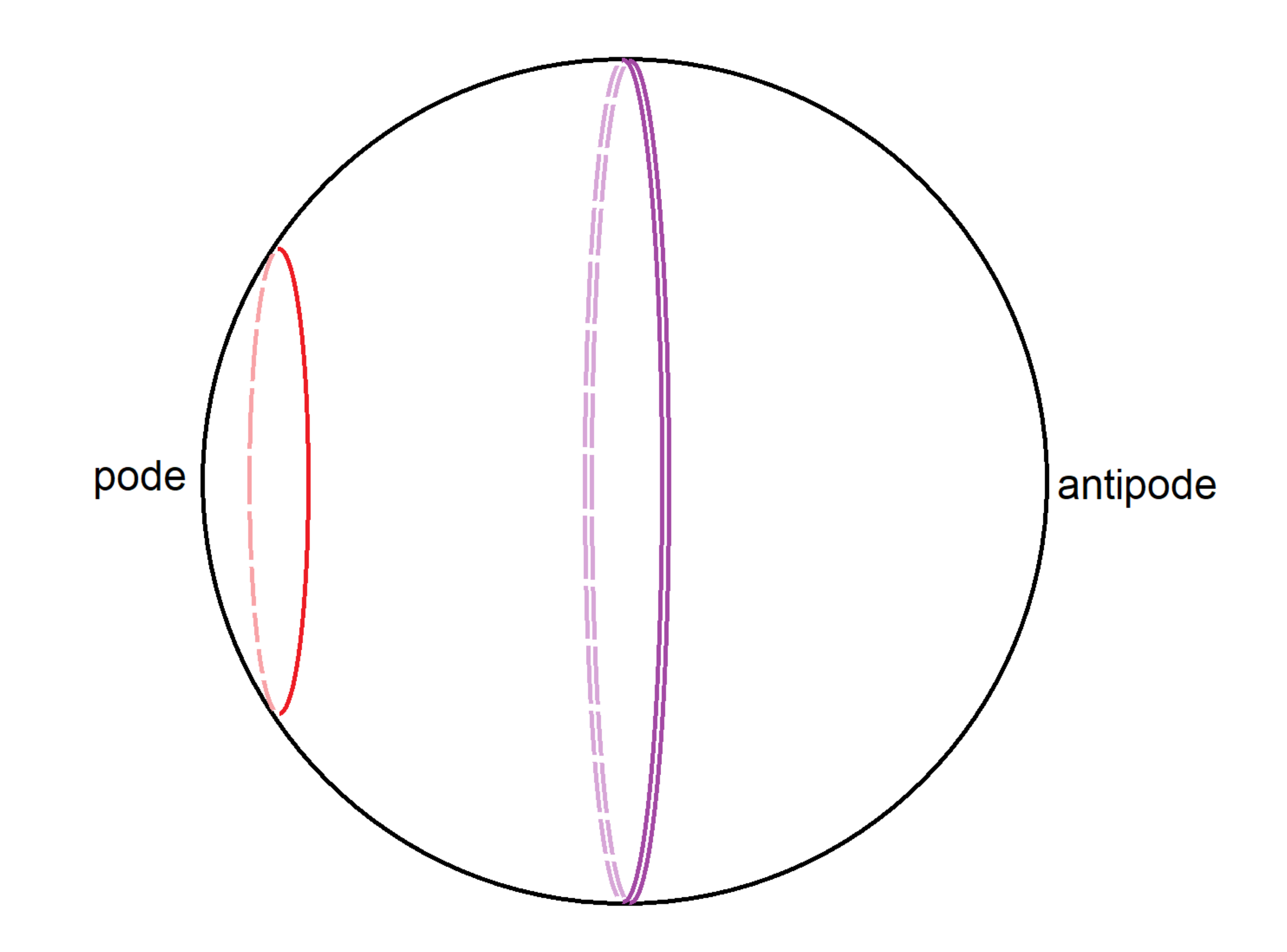}
\caption{A $t=0$ slice of dS and the stretched horizons shown as purple great circles. The red surface is homologous to the light blue horizon. It  can be shrunk to a point.}
\label{homologous1}
\end{center}
\end{figure}

\bn
This however will not work. Figure \ref{homologous1} shows the spatial slice and the adjacent pair of stretched horizons. The red curve represents a surface homologous to the pode's stretched horizon. It is obvious that that curve can be shrunk to zero, which if the above formulation were correct would imply vanishing entanglement between the pode and antipode static patches.   
\rm

\bn

\begin{figure}[H]
\begin{center}
\includegraphics[scale=.3]{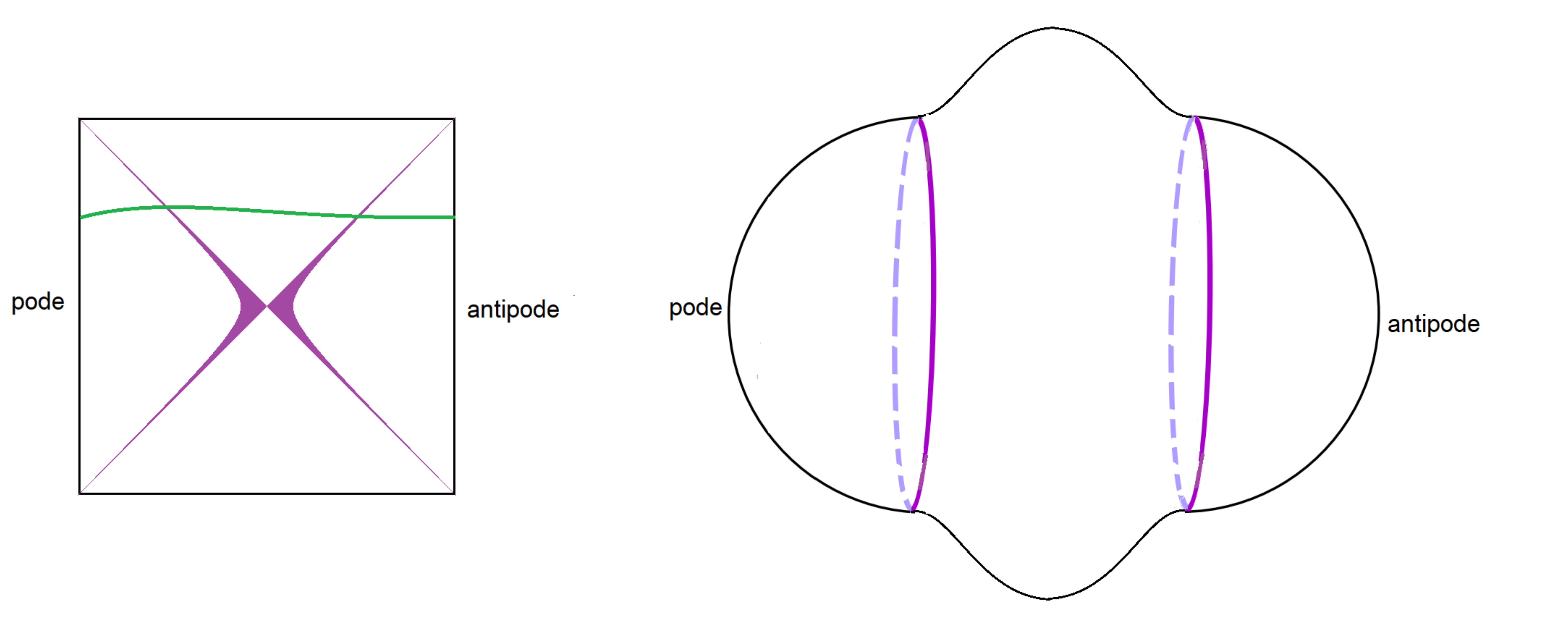}
\caption{A spatial slice through dS. In the right panel the geometry of the slice is shown.}
\label{bulge}
\end{center}
\end{figure}

To correct this problem begin by separating the two stretched horizons. This is a natural thing to do  since they will   separate after a short time, as is obvious from figure \ref{stretch}. The space-like surface shown in green in the left panel of figure \ref{bulge} passes through the separated stretched horizons.  In the right panel the geometry of the space-like slice is illustrated. 
The reason for the bulge  between the horizons is that in de Sitter space the local $2$-sphere exponentially grows as one moves behind the horizons.

Let us now reformulate a dS-improved version of the RT principle:

\bn 
\it The entanglement entropy of the pode-antipode systems is $1/4G$ times the minimal area
of a surface homologous to the stretched horizon of the pode, and \underline{lying  between  the two sets of degrees} \underline{of freedom}; 
in this case between the two stretched horizons. 
\rm

\begin{figure}[H]
\begin{center}
\includegraphics[scale=.3]{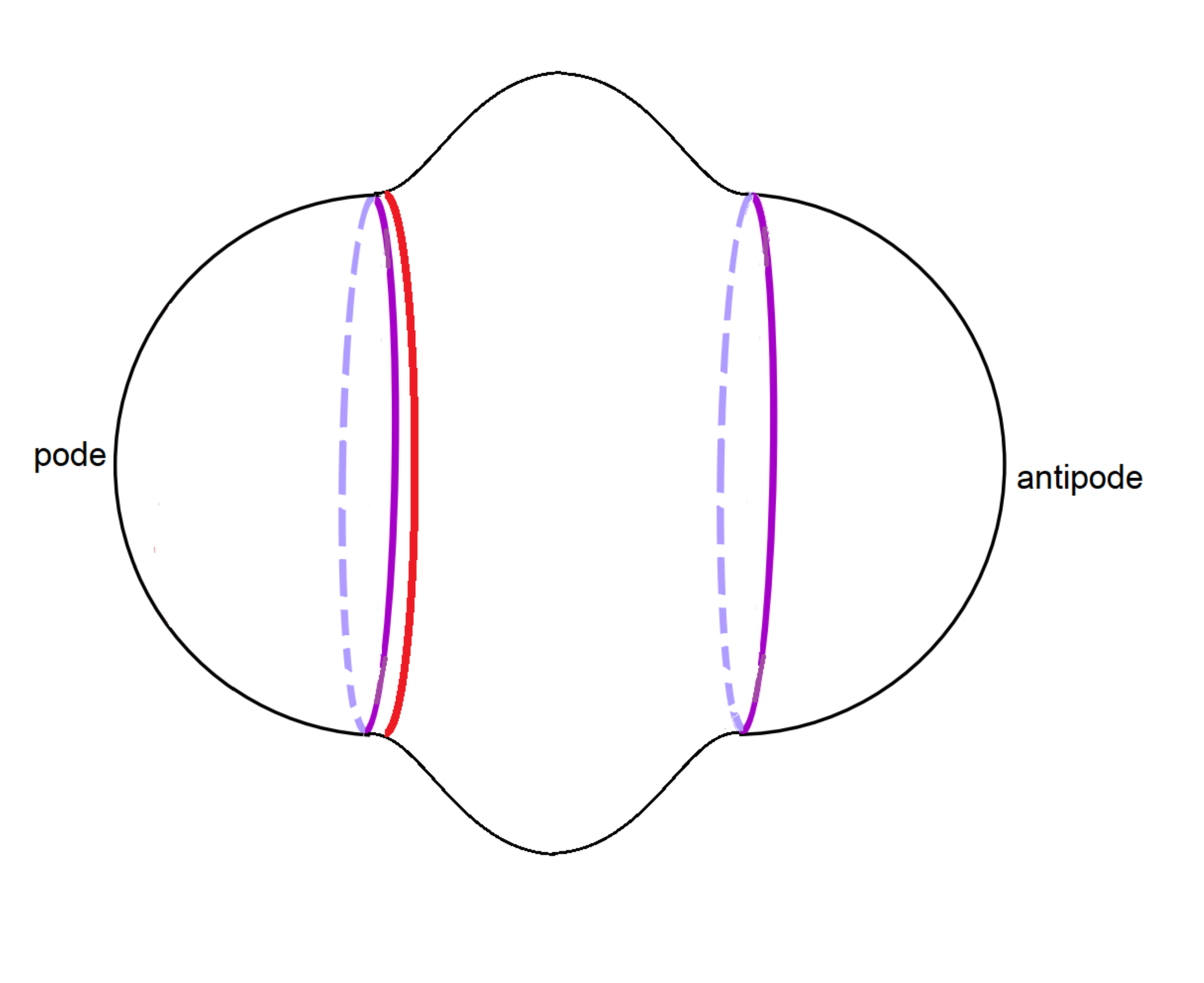}
\caption{The red curve represents the minimal  surface lying between the two stretched horizons shown in purple. There are two such surfaces, only one of which is shown,  lying right on top of the  horizons. }
\label{homologous3}
\end{center}
\end{figure}

\bn
It is evident from the geometry of the space-like slice that the minimum-area two-dimensional surface lying between the two horizons is degenerate: there are two equal minimal-area surfaces, and they lie right at the horizons.
This version of the RT principle is illustrated in figure \ref{homologous3}

\bn
Clearly the area of this de Sitter version of the RT surface is the area of the horizon.
This gives the entanglement entropy that we expect \cite{Gibbons:1977mu}, namely,
\be 
S_{ent} =\frac{\rm Horizon \ Area}{4G}.
\ee

One thing to note, is that in anti-de Sitter space the phrase ``lying between the two sets of degrees of freedom" is unnecessary. The  degrees of freedom lie at the asymptotic boundary and any minimal surface will necessarily  lie between them.

\begin{figure}[H]
\begin{center}
\includegraphics[scale=.4]{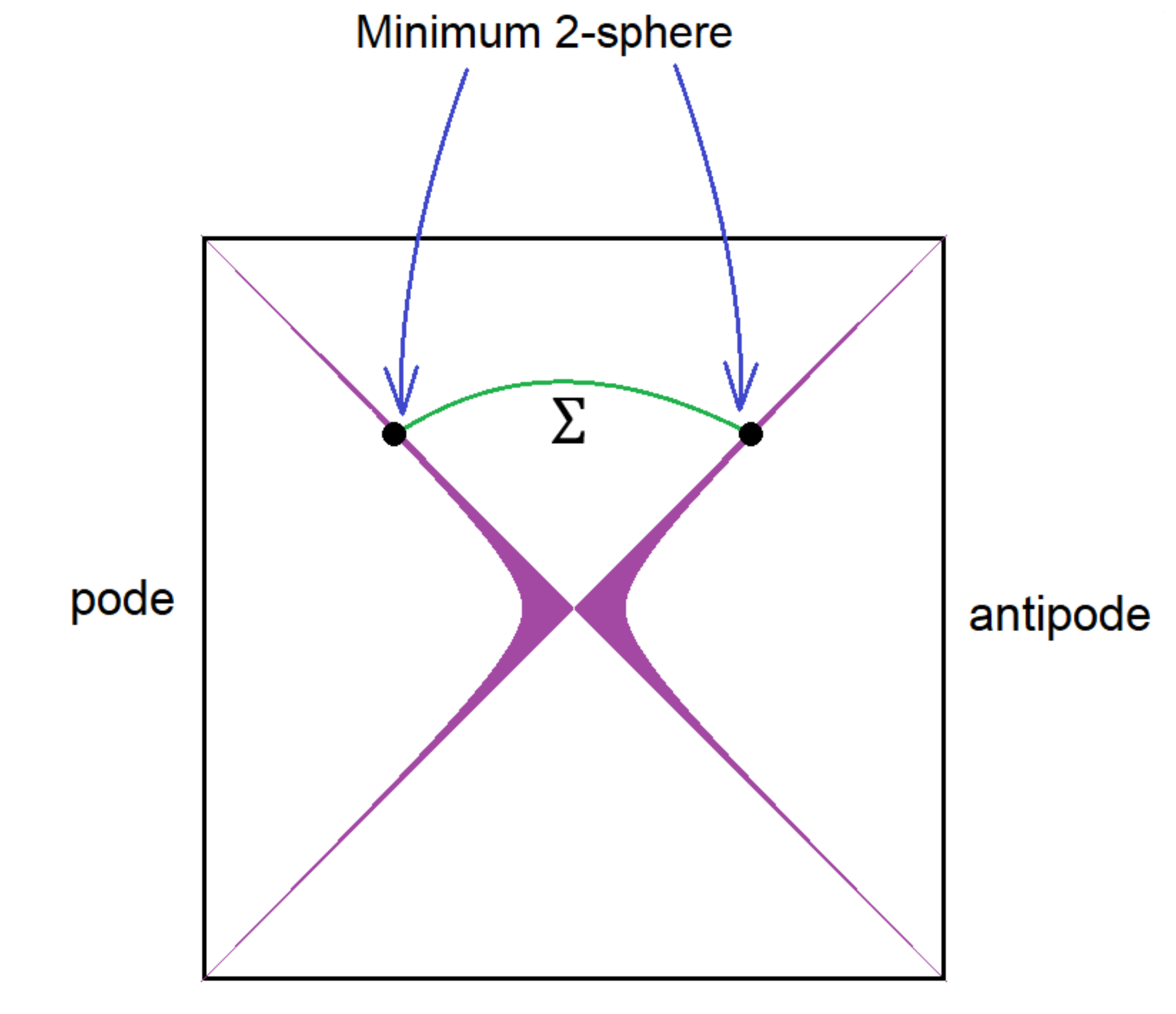}
\caption{ The black dots represent the anchoring points of a space-like surface $\Sigma$ connecting the two horizons at a particular time. The minimal two-sphere cutting $\Sigma$ lies at the anchoring points.}
\label{RTdS}
\end{center}
\end{figure}

This version of the de Sitter RT formula is sufficient for time-independent geometries. A more general HRT ``maxmin"  formulation  goes as follows:
Pick a time on the stretched horizons and anchor a three-dimensional surface $\Sigma$ connecting  the two \footnote{After version 1 of this paper appeared I became aware that Edgar Shaghoulian  had proposed a similar replacement of the AdS boundary by the dS stretched horizon, especially in the context entanglement, in a seminar at MIT last March. As in this paper Shaghoulian suggested anchoring space-like surfaces on the stretched horizon. The details of the theory appear in a recent preprint \cite{Shaghoulian:2021cef} that followed V1 of this paper.  There are some differences  with the formulation in this paper but the results are similar if not identical}. This is shown in figure \ref{RTdS}. 
Find the minimum-area two-dimensional sphere that cuts the three-dimensional surface $\Sigma$ and call its area $A_{min}(\Sigma).$ 
Now maximize $A_{min}(\Sigma)$ over all space-like $\Sigma.$ Call the resulting area $$A_{maxmin}.$$  The entanglement entropy between the pode and antipode static patches is,
\be 
S_{ent} = \frac{A_{maxmin}}{4G}.
\ee

However, in the present case, because $A_{min}(\Sigma)$ occurs at the anchoring points  the maximization of $A_{min}(\Sigma)$ is redundant: the minimum area is independent of $\Sigma.$ Later we will see an example in which this is not true.

\begin{figure}[H]
\begin{center}
\includegraphics[scale=.6]{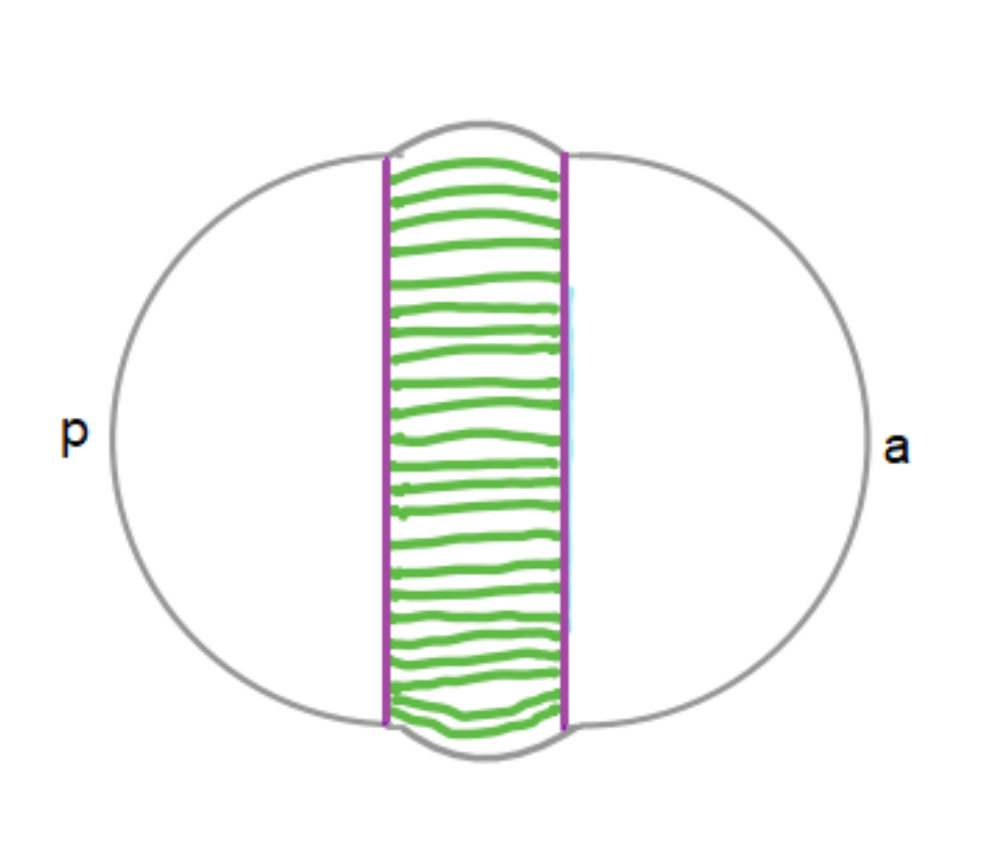}
\caption{The green lines represent bit-theads.}
\label{threads}
\end{center}
\end{figure}

All of this may be more intuitive in the bit-thread picture  \cite{Freedman:2016zud} which is illustrated in fig \ref{threads}.

\bn
We introduce a set of bit-threads ending on the two horizons. The bit-threads have a thickness or area of Planckian size. The maximum number of  bit-threads that can be squeezed through without  overlapping is the entanglement entropy. In general the minimum area surface defines the bottleneck which controls the maximum number of bit-threads. Since in the present case the
 bottleneck is at the horizon, it follows that the entanglement entropy is determined by the horizon area.

\subsection{Black Holes}

\begin{figure}[H]
\begin{center}
\includegraphics[scale=.4]{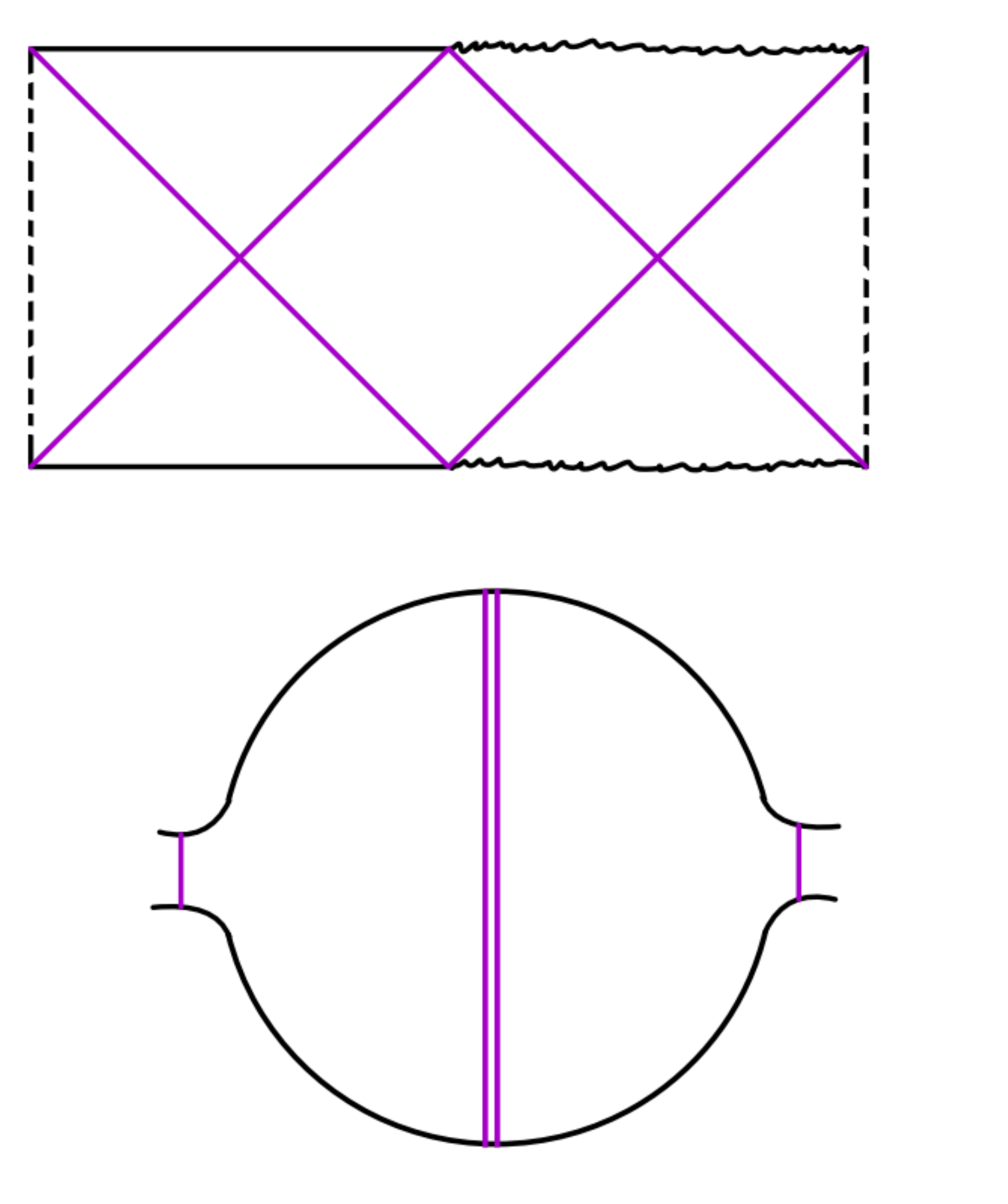}
\caption{Penrose and embedding diagrams for the \S -de Sitter black hole. The diagrams should be periodically identified along the dashed vertical lines. The two black holes at the pode and antipode are connected by an Einstein-Rosen bridge.}
\label{smallbh}
\end{center}
\end{figure}

\begin{figure}[H]
\begin{center}
\includegraphics[scale=.4]{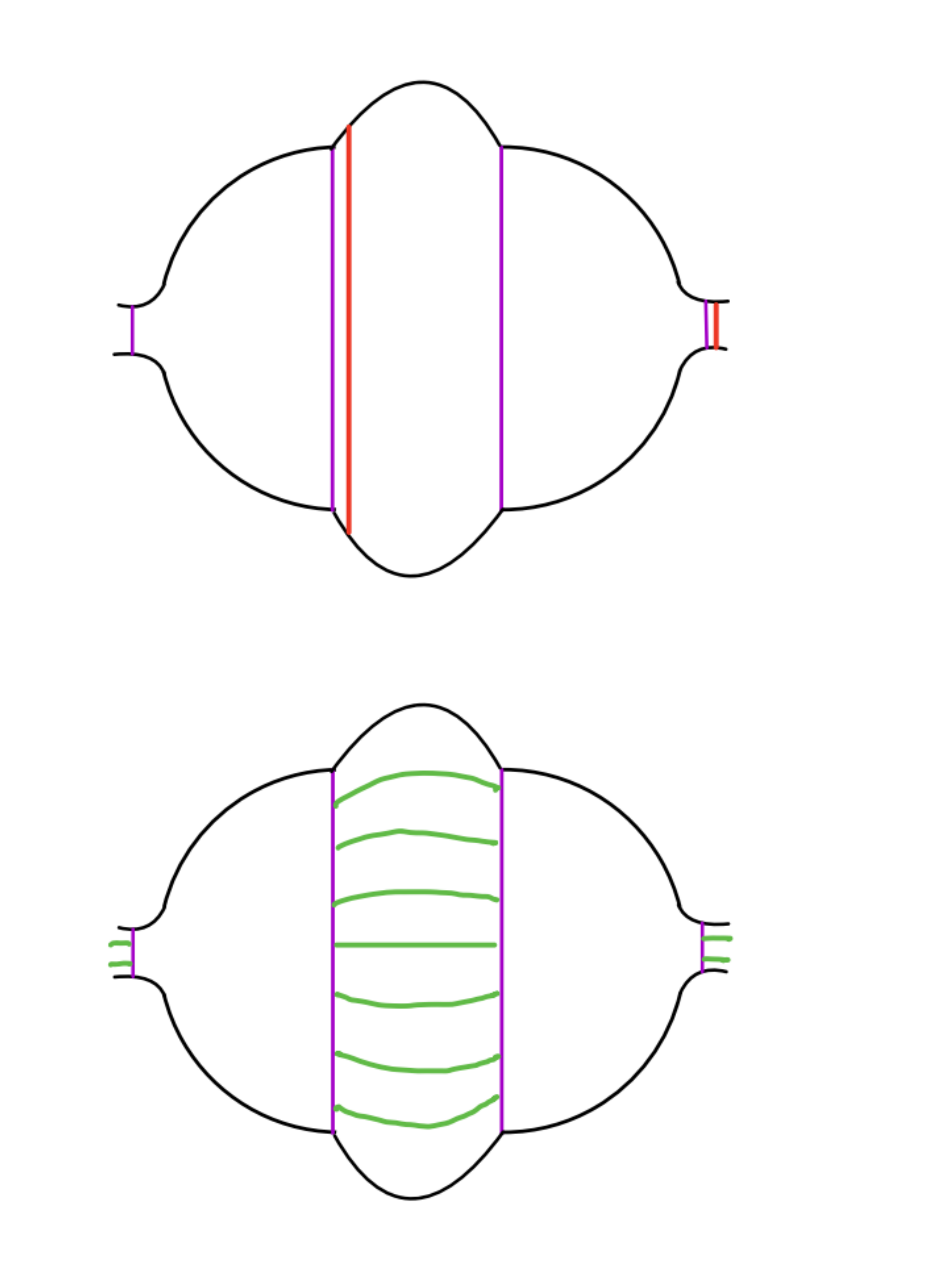}
\caption{RT surfaces and bit-thread diagram for the dS-\S \ geometry.}
\label{smallbh2}
\end{center}
\end{figure}

We come now to black holes in dS. In particular we will consider a pair of \S-de Sitter black holes located near the pode and the antipode.
The relevant Penrose and embedding diagrams are shown in figure \ref{smallbh}.
If we construct a space-like slice at a positive time it will look like figure \ref{smallbh2}, the bulge in the middle representing the exponential growth of the geometry behind the cosmic horizon. The two black holes are connected by an Einstein-Rosen bridge and are therefore entangled. 

The full horizon (generalized horizon)  of the static patch now has two components: the large cosmic horizon and the small black hole horizon. The minimal surface homologous to the generalized horizon lying between  the two  generalized horizons consists of two components shown in red and the entanglement entropy is the sum of the two areas divided by $4G.$

The bit-thread picture is especially clear. Bit theads  from the generalized horizon of one side---say the pode side---can end on the generalized horizon of the other side. The maximum number of non-overlapping bit-threads is obtained by the configuration in the lower panel of figure \ref{smallbh2}.

In this case the maxmin surface behind the smaller black hole horizon does not lie at the anchoring points but at the bifurcate horizon. That is because the black hole geometry shrinks as one moves behind the horizon.

A generalization to include bulk entanglement is possible. In fact 
the black hole contribution to the entanglement entropy may also be thought of as bulk entanglement in which the entanglement of the cosmic horizons is supplemented by the entanglement of bulk matter; namely the two entangled black holes.

\section{Hyperfast Scrambling and Quasi-normal Modes}

Now we come to the second subject of this paper: chaos and scrambling in de Sitter space.

\subsection{Scrambling}
 At first it seems obvious: if a particle is dropped from near the pode it falls to the horizon in a manner very similar to the way it would fall to the horizon of a  black hole. The momentum grows exponentially which, according to the momentum-size correspondence \cite{Susskind:2018tei} \cite{Susskind:2019ddc}, means  that the operator size also grows exponentially with the standard exponent,
\be 
size = \ e^{2\pi t / \beta}
\ee
or in terms of 
Rindler time,
\be 
size = e^{\omega}.
\ee
This implies a scrambling time (time at which the size saturates at value $S$)
\bea 
t_* &=& \frac{\beta}{2\pi} \log{S} \cr \cr
\omega_* \eq \log{S}.
\eea
Thus it would naively appear that the static patch Hamiltonian should be a  conventional fast-scrambler\footnote{This argument was put forth in \cite{Susskind:2011ap}. I now believe it is incorrect.}.

But what do we mean by size? Whose size, and how is it defined? Consider the operator that creates a particle at the pode. In the theory where size is best understood---the SYK model---size refers to the number of simple fermionic operators in the time-dependent operator that perturbs the boundary \cite{Roberts:2018mnp}. At the time the operator acts to create a fermion at the boundary its size is one. We might imagine that in dS the operator that creates a particle at the pode is also simple and of small size. This is wrong: in de Sitter holography the simple operators are located at the stretched horizon, not at the pode. The operator that creates a particle at the pode is likely to be highly complex. One sees this explicitly in the dS-Matrix theory \cite{Banks:2006rx}\cite{Susskind:2011ap}\cite{Banks:2016taq} where creating a single particle at the pode requires constraining all $N$ ``off-diagonal''  matrix \dof \ that link the cosmic horizon to the  particle. 

\subsection{Quasinormal Modes}
That raises the question: what is the bulk interpretation of the simple degrees of freedom---for example a single matrix degree of freedom---in dS holography?  The answer is clear. By analogy with AdS where simple operators describe perturbations of the boundary, in dS holography the simple operators describe disturbances of the horizon. A simple operator excites a single quantum of energy in the low angular momentum quasi-normal modes of the cosmic horizon.

\subsection{Scrambling}
The  $\log{S}$ factor in the the scrambling time in SYK
 has a simple intuitive bulk meaning as the time it takes for a single fermion to fall from the boundary, where the holographic degrees of freedom are located, to the stretched horizon \cite{Sekino:2008he}. But in  static patch holography the holographic \dof \ are already located at the stretched horizon. 
Therfore  there is no  $\log{S}$ factor in the scrambling time. The scrambling time for an initially simple operator is of the same order as the decay time of the quasinormal modes. In Rindler units that means a time of order unity,
\bea
\omega_* &\sim& 1,  \cr 
t_* &\sim& \beta = R.
\label{wstareq1}
\eea
 One could say that scrambling of the fundamental degrees of freedom in de Sitter space is ``hyperfast."  Any system that satisfies \eqref{wstareq1} I'll call a hyperfast system.

\bn
 One might object that hyperfast scrambling   violates the rigorous fast-scrambling bound of \cite{Maldacena:2015waa}. However that bound is based on an important assumption---k-locality of the Hamiltonian:

\bn
\it The meaning of k-locality: The Hamiltonian consists of a sum of monomials (in the fundamental degrees of freedom)  each of which has degree less-than-or-equal-to $k,$ where $k$ is parametrically of order unity in the limit of large entropy. In otherwords $k$ is fixed  as $R$ becomes large in Planck units.
\rm

\bn
Evidently hyperfast scrambling requires a violation of $k$ locality. One way to do this is by loosening the requirement that $k$ be independent of the overall size of the system. An example would be a system of $N$ qubits in which we retain the requirement that the Hamiltonian consists of a sum of monomials of degree less-than-or-equal-to $k,$ but allow $k$ to grow with $N$. 
 In general this will allow the system to evade the fast-scrambling bound. In section \ref{hype} we will analyze an example of this type.
 
In what follows I will reserve the term \kl \  strictly for the case in which $k$
is fixed and independent of $N.$

\subsection{Scrambling and Thermalization}
Scrambling and thermalization are two different phenomena. Scrambling is diagnosed by  out-of-time-order four-point functions while thermalization is diagnosed by the conventional two-point function or more simply the decay rate of a simple perturbation. For example the decay time of quasinormal modes for black holes is order $\beta,$ the inverse temperature, while the scrambling time is of order $\beta \log{S}.$   

Nevertheless I am arguing  that  de Sitter space is a hyperfast system meaning that the two time-scales---scrambling time and quasinormal mode decay time---are of the same order,
\be 
t_{thermal} \sim t_* \sim \beta.
\label{hyperfast}
\ee
In section \ref{hype} I will give a quantitative argument for this behavior in a certain limit of  qubit models including SYK.

\section{Hyperfast Complexity Growth}

Hyperfast scrambling has implications for complexity growth. For \kl \ Hamiltonians complexity growth is bounded and cannot exceed  linear growth with a coefficient proportional to the product of temperature and entropy. Typically this is saturated for black holes. Equivalently, for black holes complexity grows with the  time like,
\be 
\CC \sim STt =    S \omega
\label{CsimSw}
\ee

But if the Hamiltonian itself is complex---for example if it contains terms of very high order in the elementary degrees of freedom---then one might expect complexity growth to be hyperfast. This, in turn,  should manifest itself in the space-time geometry through the holographic complexity dualities. We will see that  hyperfast complexity growth is closely related to the key property of de Sitter space: its exponential growth.

\subsection{Maximal Slices in dS} \label{maxslice}
Let us consider generalizations of holographic complexity to de Sitter space. First the CV proposal. Instead of  anchoring spatial slices  on the boundaries (which do not exist) we anchor them on the horizons.  Then, subject to the  anchoring conditions, we determine the maximal space-like surface that stretch between the two horizons. The holographic complexity conjecture is that the complexity is proportional to the volume of those slices.

\begin{figure}[H]
\begin{center}
\includegraphics[scale=.5]{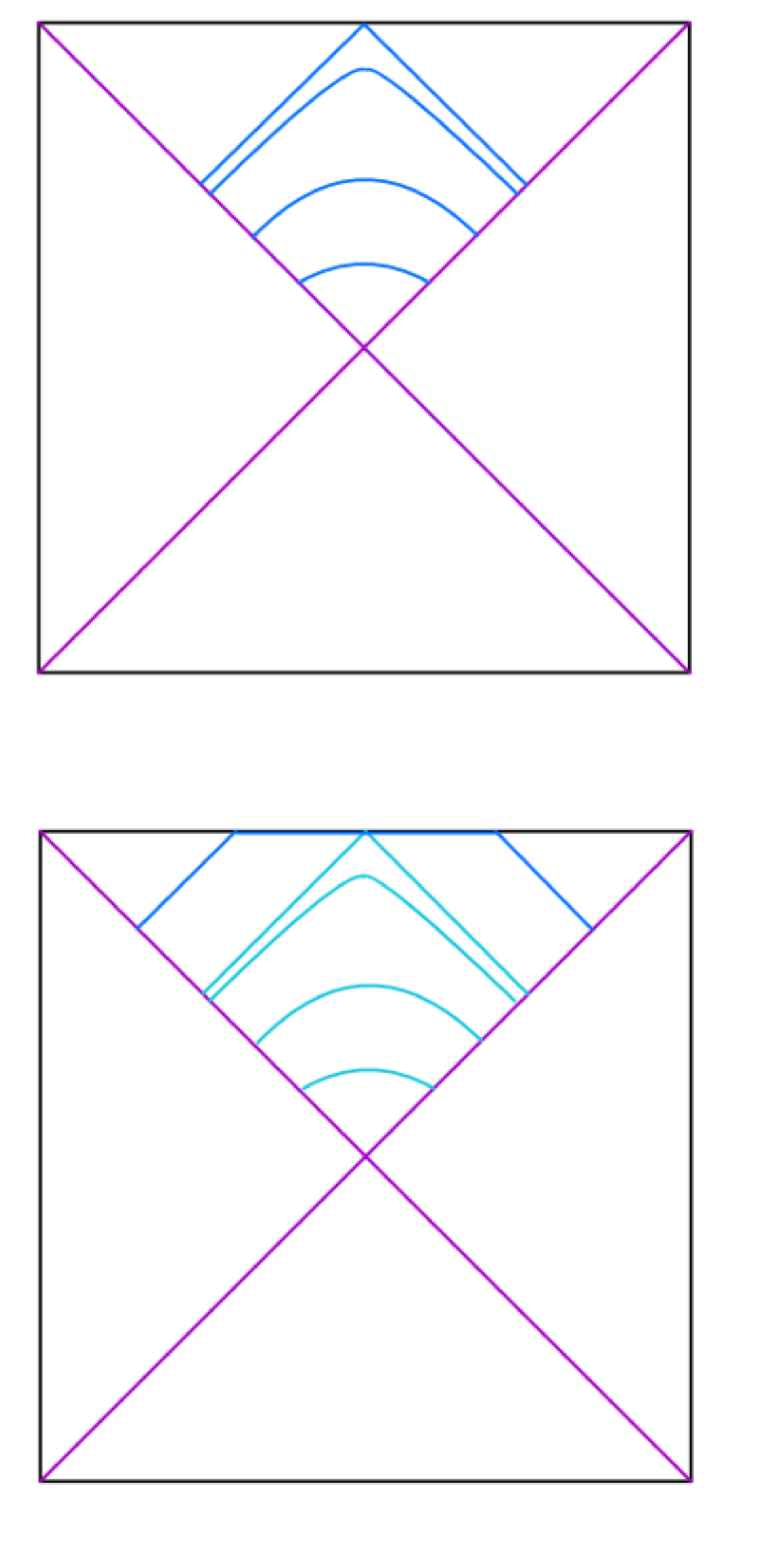}
\caption{The top panel shows the early smooth maximal slices bending upward due to the expanding local two-spheres. At some point the maximal surfaces intersect future infinity.  The bottom panel shows the maximum slices, assuming an appropriate cutoff, at later times. }
\label{CV}
\end{center}
\end{figure}

For early times the maximal slices are smooth but relative to AdS they bulge up into the interior region as shown in figure \ref{CV}. 

\bn
The reason is that the geometry of de Sitter space exponentially grows behind the horizon. Therefore, in order to capitalize on that growth, the maximal surfaces bend upward toward the future boundary. Let us consider the midpoint of the maximal surface and assume that it is a proper time $\tau$ from the bifurcate horizon. At the midpoint the area of the local two-sphere grows like $\exp{\frac{\tau}{R}}.$  This has the effect that  the volume of the maximal slice also grows exponentially.

As figure \ref{CV} shows, a point quickly occurs at which the surfaces become light-like and  reach the future boundary. This happens after a time of order $\beta$ or a Rindler time of order unity.

Beyond this point the classical rate of volume growth  becomes infinite. Since the true complexity is limited to be exponential in the entropy, I take this to mean that the rate of complexity growth  becomes exponentially large once the anchoring points have moved a short distance along the stretched horizons. A reasonable interpretation is that for a hyperfast system, after a Rindler time  of order unity the complexity grows linearly with time, but with a rate that is non-polynomially large in $S,$ 
%
%
\bea 
\CC(\omega) &\sim&    \exp{\lf\sqrt[a]{S}\rg}  \times \omega  \cr
1<&a&<\infty,
\label{nonpoly}
\eea
instead of as in \eqref{CsimSw}.
Later in section \ref{sykds} we will see that this is exactly  what is predicted by quantum circuits in the hyperfast regime.

\subsection{Inflation: The Dual of Hyperfast Complexity Growth}

The  growth of maximal surfaces is a geometric feature whose dual is complexity growth. 
Hyperfast growth of maximal surfaces, as in figure \ref{CV}   reflects the inflationary growth of the interior of de Sitter space. (The same things are true for the complexity-action duality.)
This  leads to the following conjectured duality:

\bn
\it
The property of hyperfast complexity growth is the holographic dual of the exponential growth of de Sitter space behind the horizon of the static patch.  
\rm

\bn

What does it take for complexity to grow in a hyperfast way? A necessary condition is that the Hamiltonian not be \kl \ (unless $k$ is allowed to grow with $N$).

\section{SYK and dS}\label{sykds}

I now want to conjecture that a particular limit of a well-known system displays features of de Sitter space: 

\subsection{A Radical Conjecture}

\it 
\bn
The SYK system, and other similar qubit systems, in the  limit of high temperature (but finite tomperature), with $q$ scaling as a power of $N,$
\bea 
T&\to& \infty \cr
q &\sim& N^{p} \cr
0&<&p<1,
\label{hyperlimit}
\eea
are hyperfast and have features that strongly resemble  de Sitter space. 
\rm

In particular systems defined by \eqref{hyperlimit}  are  hyperfast scramblers with the scrambling  and thermalization times  of the same order, as in equation \eqref{hyperfast}. Both time scales are set by the fundamental energy scale of SYK, namely the parameter $\CJ.$ 
\be 
t_{thermal} \sim t_* \sim {\CJ}^{-1},
\label{TtsimJ}
\ee
where $\CJ$ is the usual   SYK energy scale that controls the magnitude of the $q$-local terms in the Hamiltonian. 

Most  significantly, the SYK system in the   limit \eqref{hyperlimit} exhibits hyperfast complexity growth, the importance of which I explained in section \ref{maxslice}. In section \ref{hype}  an
explicit calculation will confirm this\footnote{I thank Adam Brown for sharing his calculations of  complexity growth in SYK at large $q.$ }.

\bn

\subsection{The Boundary Becomes the Horizon}
\begin{figure}[H]
\begin{center}
\includegraphics[scale=.4]{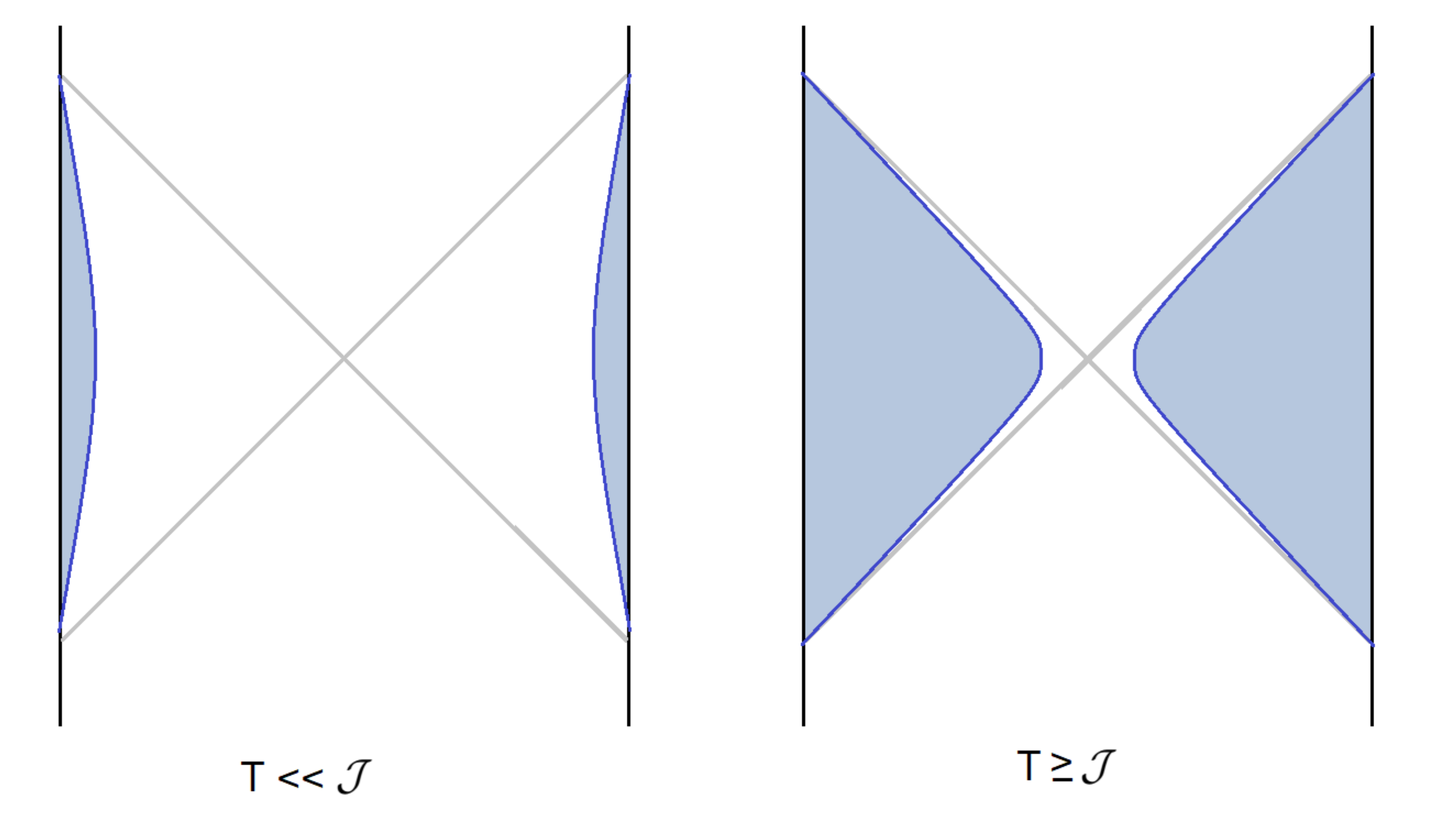}
\caption{Penrose diagram for the JT bulk dual of SYK.  The dynamical Schwarzian boundary bends inward toward the horizon. At low temperature shown in the left panel the boundary bends in very little. At high $T$ it bends inward and almost hugs the horizon.}
\label{JTDS}
\end{center}
\end{figure}

SYK in this parameter regime has other similarities with de Sitter holography. Consider the Penrose diagram for the bulk dual of SYK; namely JT gravity. The geometry for low temperature is  shown in  the left panel of figure \ref{JTDS}.
As is well known, the boundary of the geometry  bends inward toward the horizons by an amount that depends on the temperature \cite{Maldacena:2016upp}. For very low $T$ the boundaries bend in very little and the distance between them at $t=0$ is large as in the left panel. But as the temperature increases the boundaries move closer to the horizon and if one follows the formulas to $T \to \infty$  (or tomperature to $\CJ$) they get within a cutoff distance 
${\CJ}^{-1}$ of the horizons as in the right panel. In the high temperature limit the boundary merges with the stretched horizon and the holographic degrees of freedom cannot be distinguished from horizon modes. To put it another way, the boundary degrees of freedom become the quasinormal horizon modes.

If we combine this picture with the conjecture that for $q\sim N^{p}$ complexity growth (see next section \ref{hype}) is hyperfast then it follows that the geometry between the horizons grows rapidly, just as one might expect for de Sitter space.


All in all, I think it is possible that in the limit  $T\to \infty, \ \CT \to \CJ,$ and   $ \ q\sim N^{p},$ the SYK system may have more in common with  de Sitter space than with a black hole. But two things are missing: an argument 
that there is an emergent space-time in the grey shaded region of the right panel of figure \ref{JTDS}, i.e., geometry near the pode and antipode; and evidence that the model has the symmetries of de Sitter space. 

As I argued earlier, the degrees of freedom near the pode region should be constructed from somewhat complex combinations of the horizon degrees of freedom
and might not be easy to see.  As for the symmetries I don't have anything to say right now.

  \subsection{Hyperfast Scrambling in SYK}\label{hype}
  
  This rest of this  section is based on calculations that were explained to me by Adam Brown, Steve Shenker, Douglas Stanford, and Zhenbin Yang. The detailed calculation in the Brownian\footnote{As in Brownian motion.}   ``epidemic model" was done by Adam Brown. It  applies to the SYK model, but also more general qubit models in the limit of large temperature and large k-locality parameter which in SYK is called $q$.
  
  At high temperature it is sufficient to consider so-called Brownian models. We begin with an epidemic model for operator growth. The rule is that in each step of a quantum circuit $\lf N/q \rg$ random q-local gates act on the $N$ qubits. Each qubit is represented in one gate and no qubit interacts in more than one gate. 

At the start one qubit is ``infected." After $n$ steps the probability that  any qubit is infected is $P(n).$  According to Brown, $P(n)$ satisfies the recursion relation,
\be  
P(n+1) = P(n) + \big[ 1-P(n) \big]  \big[  1 - \big( 1-P(n)\big)^{q-1}    \big].
\label{discp}
\ee

In the continuum Brownian limit this becomes,
\be 
\frac{dP}{dn} = \big( 1-P \big) \lf(  1 - (1-P)^{q-1}    \rg.
\label{cntinp}
\ee

\begin{figure}[H]
\begin{center}
\includegraphics[scale=.6]{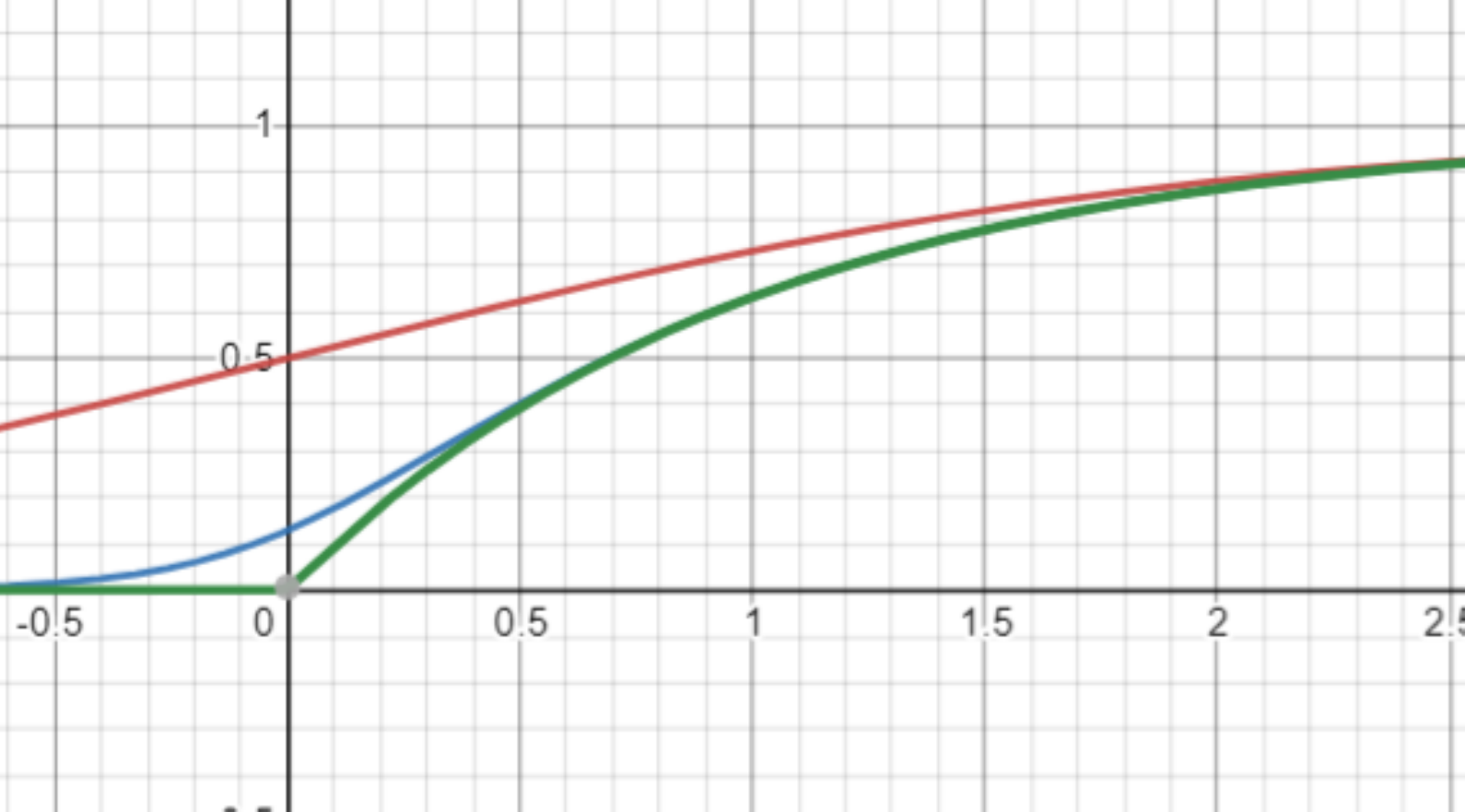}
\caption{The function in \eqref{solut} for $q=2$ (red), $q=6$ (blue), and  
$q=100$ (green).  By the time we get to $q=100$ the curve has become indistinguishable (for $n>0$) from the late-time behavior in \eqref{earlylate}.
The very early exponentially growing  region collapses to zero as $q$ increases. }
\label{graph}
\end{center}
\end{figure}

Integrating \eqref{cntinp} gives,
\be 
P(n) = 1 - \big( 1+e^{(q-1)n}   \big)^{\frac{-1}{q-1}}
\label{solut}
\ee
shown in figure \ref{graph}.

The solution involves two exponential behaviors:
\bea 
P(n) &=&  \frac{1}{N} e^{(q-1)n} \ \ \ \ \ \ \ \  {\rm early \ time} \cr \cr
P(n) &=& 1-  e^{-n}   \ \ \ \ \ \ \  {\rm late \ time}
\label{earlylate}
\eea

The early behavior is the usual limiting Lyapunov behavior associated with the scrambling bound \cite{Maldacena:2015waa}. This determines the relation between $n$ and the Rindler time for the usual SYK theory,
\be 
\omega = (q-1)n.
\ee
The scrambling time is determined by $P(n)\to 1,$ 
\be 
n_* = \frac{1}{q-1} \log{N}.
\ee

But analyzing the solution \eqref{solut} we see that as $q$ increases the range of time over which the Lyapunov behavior is relevant shrinks to zero with large $q.$ This is easily seen in figure \ref{graph}. When $q$ is extremely large the scrambling time $n_*$ is determined by the late-time exponent and we find,
\be 
n_* =1.
\ee
This applies whenever $q\sim N^p.$

On the other hand the thermal time  (the time-constant for the decay of the two-point function) in the limit $q\sim N^p$ can also be calculated from the Brownian model.  One finds that the thermalization time is proportional to the scrambling time\footnote{I thank Zhenbin Yang for this observation.},
\be 
\omega_{therm} = \frac{\omega_*}{2}.
\ee
We may interpret this to be dual to the statement that the scrambling time and the decay time for quasinormal modes are equal apart from a factor of order unity. This, as I explained, is what one expects if the holographic degrees of freedom are located on the stretched horizon of the  de Sitter static patch.

The thermal time scale is also the decay time for quasinormal modes, and in Rindler units it is of order unity. This observation allows us to identify the circuit time $n$ with Rindler time up to a constant of order unity,
\be 
\frac{n}{\omega} \sim 1.
\ee

\subsection{Hyperfast Complexity Growth in SYK}

In an attempt to place bounds on complexity growth in quantum circuits Nielsen 
\cite{Nielsen} introduced a right-invariant ``complexity geometry" on the group $SU(2^N).$ The geometry is characterized by certain penalty factors\footnote{The notation in this section is that of Brown and Susskind \cite{Brown:2017jil}} $\CI(w)$ which Nielsen took to be exponentially large or even infinite, in all directions other than those associated with operators involving $k$ or fewer qubits,
\bea 
\CI(w) &=&1  \ \ \ \ \ \ \ \ \  w\leq k \cr \cr
\CI(w) &\geq& 4^N  \ \ \ \ \ \ \ \ \  w>k
\eea
The geometries with infinite penalty factors are called sub-Riemannian and have pathological properties such as infinite sectional curvatures. The Nielsen and sub-Riemannian geometries are difficult to work with. Technically the reason is that the cut-locus is very close to the origin. Nevertheless \cite{Nielsen} was able to derive some interesting bounds.

Nielsen in his work did not suggest that geodesic length (in the
complexity geometry) was an actual definition of complexity. That
suggestion was made by Brown and Susskind \cite{Brown:2017jil}, who argued for a
geometry with a much more moderate schedule of penalty factors,
\be 
\CI(w) \sim 4^w.
\ee

In this section I will assume results from a soon-to-be published
paper by Adam Brown, Henry Lin, Michael Freedman, and myself in which we argue that there is a universality of complexity geometries. The conjecture of universality is the following:

\begin{itemize}
\item  The complexity growth for the sub-Riemannian, Nielsen, and
Brown-Susskind geometries are approximately the same. Moreover the same can be said for any schedule lying between subRiemannian and Brown-Susskind.

\item In generic directions the cut locus moves from being very close
to the origin for Nielsen geometry, to being exponentially far
from the origin for the Brown-Susskind geometry.
\end{itemize}

The details are not important for us in this paper. Two things
are important: first that the great distance of the cut locus from the
origin makes it easy to use the geometry in order to estimate how
complexity grows, for very long times. Second, the Brown-Susskind
schedule implies that at large $q$, after a short transient time the
complexity growth for the SYK system satisfies,
\be 
\CC(t) = 2^q \epsilon Nt
\ee
where $\epsilon$ is the variance in the energy per degree of freedom,
\be 
\langle H^2\rangle =N\epsilon^2.
\ee

Ordinarily for a system at high temperature $\epsilon$ would grow with
temperature, but for reason discussed in section \ref{tom}  we identify it
with the tomperature which remains finite as $T\to \infty.$

Thus, the complexity as a function of time is,
\be 
\CC(t) = 2^q \CT S t,
\ee
where I’ve used the fact that the entropy is $\sim N.$ Finally the 
dimensionless time $\omega$ is defined by,
$\omega =\CT t.$
The complexity takes the form,
\be 
\CC(\omega) =2^q S \omega.
\ee

Next let us go to the hyperfast limit, $q = N^p$,
\be 
\CC(\omega)=\lf  2^{N^p} \omega. \rg
\label{531}
\ee
This is to be compared with the ordinary complexity growth for
an ordinary fast-scrambler,
\be
\CC(\omega) \sim N \omega. 
\label{532}
\ee
The rate of growth in \eqref{531}  is exponentially larger than in \eqref{532}
in exactly the way described in \eqref{nonpoly} (with $a = 1/p$). In the large
$N$ limit it is infinitely faster than \eqref{532}, as required by the lower
panel of figure \ref{CV}. The only bulk interpretation that I know of is
that the geometry between the horizons grows exponentially, just
as it does in de Sitter space.

The behavior in \eqref{531} breaks down for short times $t \leq \CT^{-1} $
and
for exponential times when the complexity has reached its maximum value and must stop growing.
One can use  \eqref{531} to compute how long it takes to reach maximum complexity. The maximum complexity of a $N$-qubit system  evolving under the action of a time-independent Hamiltonian is $2^N$.
The time that it takes to reach that value is,
\be 
\omega \sim 2^{N-N^p} \sim 2^N.
\ee
Thus even with hyperfast growth the complexity still takes time
of order $2^N$ to reach its maximum, after which classical GR must
break down for the global space-time geometry.

This completes the argument that SYK (and other qubit models)
at high temperature and $q = N^p$
are hyperfast. I have argued that
the bulk dual of hyperfastness is:
\begin{enumerate}
\item  A scrambling time equal (within a factor of $2$) to the decay
time of quasinormal modes.
\item Exceptionally rapid growth of volume between the horizons,
dual to the hyperfast growth of complexity.
\end{enumerate}

Both of these are \it inconsistent \rm with the behavior of black holes,
but are distinctive features of de Sitter space.
If correct, the conjectures of this paper open a new direction of
research in which the tools of quantum information theory can be
applied to cosmological space-times.

\section*{Acknowledgements}
I thank Adam Brown, Edgar Shaghoulian, Steve Shenker, Douglas Stanford, and Zhenbin Yang for crucial discussions on the material in section \ref{hype}.

\bn
LS was supported in part by NSF grant PHY-1720397.

\section*{Notes Added}
\begin{enumerate}
\item The special case of $q=N^{p}$  with $p=1/2$ is called the double-scaling limit of SYK and was studied more fully  in \cite{Cotler:2016fpe}\cite{Berkooz:2020uly}.  I thank Henry Lin for pointing this out.

\item
I can't recall who it was, but after this paper was circulated someone suggested a similarity with mechanisms discussed by Silverstein and collaborators \cite{Lewkowycz:2019xse}\cite{Shyam:2021ciy} in their discussion of the static patch. Consider going from
 low temperature  at fixed $q$  (the conformal limit)  to the limit $T\to \infty, \ q\to N^p$ in two steps. First, keeping $q$ fixed we increase $T$ which has the effect of pulling in the boundaries as in figure \ref{JTDS}, but without driving the system into the hyperfast regime. This pushes the theory from a low temperature near-extremal black hole to a higher temperature non-extremal black hole, but not toward de Sitter.

Then we let $q$ increase. That is what brings the system from black hole behavior to a de Sitter-like phase.

Compare this  with Silverstein and collaborators who first deform a holographic CFT by a $\bar{T} T$ deformation. If I understand correctly, this has the effect of pulling in the boundaries in much the same way as a cutoff; again illustrated by figure \ref{JTDS}. In a certain sense it is also increases the temperature, or more exactly the ratio of the temperature to the cutoff scale of the CFT. Of course at the end there is not much left of the CFT. 

If the $\bar{T} T$ deformation is anything like a conventional cutoff this makes the theory spatially non-local but does not affect the k-locality of the couplings. (The couplings in an ordinary gauge theory are at most quartic which means the theory is 4-local.)

They then follow this operation by what they call a $\Lambda$ deformation, which, it is argued,  drives the system to a de Sitter phase.  If it does produce de Sitter space with a inflating interior, then by the arguments  in this paper it should be possible to show that the $\Lambda$ deformation leads to a breakdown of k-locality, possibly by the introduction increasingly high powers of the fields, and to hyperfast scrambling.

\item  Herman Verlinde has called to my attention a number of talks in which he also conjectured a relation between SYK in the double scaling region and de Sitter space \cite{Verlinde}. See also Akash Goel and Herman Verlinde, in preparation.

\end{enumerate}

\end{document}